\documentclass[aps,prl,showpacs,amsmath,amssymb,longbibliography, notitlepage,superscriptaddress, twocolumn]{revtex4-2} 
\usepackage[dvipsnames]{xcolor}
\usepackage{floatrow}
\usepackage{soul}
\usepackage{siunitx}
\usepackage{physics}
\soulregister\cite7
\soulregister\ref7
\usepackage{graphicx}
\usepackage{placeins}
\definecolor{darkblue}{rgb}{0,0,0.6}
\definecolor{darkred}{rgb}{0.6,0,0}
\usepackage[colorlinks=true,urlcolor=darkblue,citecolor=darkblue, linkcolor=darkred,hyperfootnotes=false]{hyperref}
\usepackage[color=yellow, size=footnotesize]{todonotes}
\RequirePackage{lettrine} 

\setul{}{1.2pt}

\makeatletter
\DeclareFontFamily{U}{tipa}{}
\DeclareFontShape{U}{tipa}{m}{n}{<->tipa10}{}
\newcommand{\arc@char}{{\usefont{U}{tipa}{m}{n}\symbol{62}}}%
\newcommand{\arc}[1]{\mathpalette\arc@arc{#1}}
\newcommand{\arc@arc}[2]{%
  \sbox0{$\m@th#1#2$}%
  \vbox{
    \hbox{\resizebox{\wd0}{\height}{\arc@char}}
    \nointerlineskip
    \box0
  }%
}
\makeatother

\begin{document}

\title{A wrinkled cylindrical shell as a tunable locking material}

\author{Pan Dong}
\affiliation{Department of Physics, Syracuse University, Syracuse, NY 13244}
\affiliation{BioInspired Syracuse: Institute for Material and Living Systems, Syracuse University, Syracuse, NY 13244}
\author{Mengfei He}
\affiliation{Department of Physics, Syracuse University, Syracuse, NY 13244}
\affiliation{BioInspired Syracuse: Institute for Material and Living Systems, Syracuse University, Syracuse, NY 13244}
\author{Nathan C. Keim}
\email{keim@psu.edu}
\affiliation{Department of Physics, Pennsylvania State University, University Park, PA 16802}
\author{Joseph D. Paulsen}
\email{jdpaulse@syr.edu}
\affiliation{Department of Physics, Syracuse University, Syracuse, NY 13244}
\affiliation{BioInspired Syracuse: Institute for Material and Living Systems, Syracuse University, Syracuse, NY 13244}

\begin{abstract}
A buckled sheet offers a reservoir of material that can be unfurled at a later time. For sufficiently thin yet stiff materials, this geometric process has a striking mechanical feature: when the slack runs out, the material locks to further extension. Here we establish a simple route to a tunable locking material -- a system with an interval where it is freely deformable under a given deformation mode, and where the endpoints of this interval can be changed continuously over a wide range. We demonstrate this type of mechanical response in a thin cylindrical shell subjected to axial twist and compression, and we rationalize our results with a simple geometric model. 
\end{abstract}

\maketitle

\begin{figure}[t!]
	\includegraphics[width=1.0\textwidth]{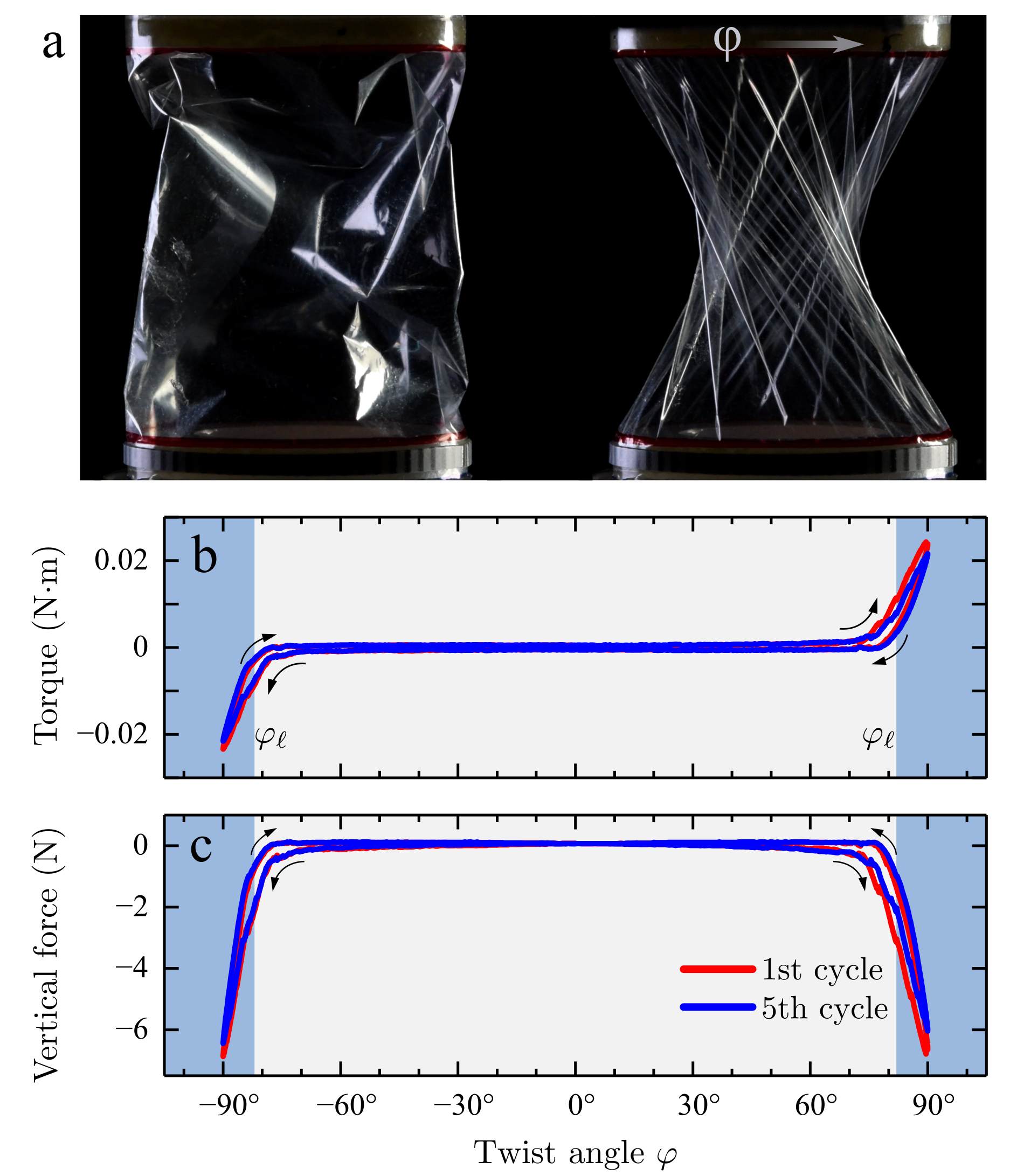}
	\caption{
	\textbf{Morphology and mechanics of a twisted cylindrical shell.}
	(a) A 6 $\mu$m thick shell of height $H=26.9$ mm and radius $R=9.5$ mm is mounted between parallel plates, compressed axially 
 and then twisted. At a threshold twist angle, the shell transitions from a disordered buckled state to an ordered wrinkled state. Left: $\varphi=0$. Right: $\varphi=90^\circ$.
	(b,c) Torque and vertical force on the top plate, as this cylinder is twisted 
	cyclically between $\pm 90^{\circ}$ at constant separation ($h=22.2$ mm). 
	The 
	curves are reproducible over multiple cycles.
	Coinciding with the transition to ordered wrinkles, the data rapidly increase in magnitude at the ``locking angle'' $\varphi_\ell$. Our model predicts $\varphi_\ell = 81^{\circ}$ for this shell (see Eq.~\ref{eq:phi_l}). 
	}
	\label{fig:force_torque}
\end{figure}

When a thin material is compressed, small-scale wrinkles and folds may form to collect excess length. 
Looking at this process in reverse, buckled microstructures can be thought of as a reservoir of material that can be deployed at a later time. 
This strategy is exploited in designed structures   
ranging from the common umbrella to inflatable satellites \cite{Lennon05,Paulsen19}. It is also harnessed in nature, for instance in the capture thread of the orb-weaving spider \cite{Elettro16,Grandgeorge18} and in rabbit mesentery \cite{Fung67}. 
The process of buckling and deploying material length  is rich in its geometric aspects \cite{Tobasco22}, and it also has a striking mechanical feature: 
a wrinkled sheet has approximately no resistance to extension until it becomes taut, at which point the force to stretch the system further rapidly increases (Fig.~\ref{fig:force_torque}c). 
This mechanical response has been idealized in a theory of so-called ``locking materials'' \cite{Prager57,Prager69}.

Here we establish a simple route to a \textit{tunable} locking material -- a system with an interval where it is freely deformable, and where the endpoints of this interval can be changed continuously over a wide range. We demonstrate this type of mechanical response in a thin cylindrical shell subjected to axial twist and compression. Our measurements show that this system is soft to twisting up to a threshold ``locking angle’’ $\varphi_\ell$. By adapting the basic physical picture of tension field theory \cite{Wagner29,Pipkin86,Mansfield89,Steigmann90} into simple length-preserving arguments, we predict the locking angle as a function of a small set of geometric parameters. We predict a universal phase boundary between the soft deformations facilitated by buckling and the stiff response where the system becomes taut, in agreement with our experiments. These results provide a prototypical example of a locking position that can be tuned \emph{in situ}, which could find use in applications where a reprogrammable mechanical response is desired.

The phenomenon we study is shown in Fig.~\ref{fig:force_torque}. Here, a rectangular mylar film (thickness $t=6$ $\mu$m, Young’s modulus $E=3.4$ GPa) has been curved into a cylindrical shape by gluing two of its edges to rigid rings of radius $R=9.5$ mm, which are mounted in a rheometer (Anton Paar MCR 302) for mechanical testing. Undeformed, the cylinder strongly resists extension, but will readily buckle and wrinkle under compression. Starting from an initial height of $H=26.9$ mm, the cylinder is axially compressed to a height $h=22.2$ mm so that it buckles into a loosely crumpled configuration (Fig.~\ref{fig:force_torque}a). Then, we slowly rotate the top ring. When the magnitude of the rotation angle $\varphi$ reaches approximately $81^{\circ}$, we witness two striking events at once: (i) the complex crumpled state gives way to regular wrinkles, and (ii) the magnitude of the torque on the top ring begins to increase dramatically (Fig.~\ref{fig:force_torque}b). 
Performing this deformation cyclically reveals that these mechanical measurements and morphological transitions are repeatable. 

The normal force on the top ring during the experiment also shows a dramatic increase in magnitude at the same angle $\varphi_\ell$, as we show in Fig~\ref{fig:force_torque}c. The wrinkled morphology of Fig.~\ref{fig:force_torque}a shows why the two signals are linked: stresses are transmitted between the bottom and top rings along the wrinkle crests and troughs, which are tilted with respect to the vertical axis. Thus, we may use the normal force as a secondary signal to measure $\varphi_\ell$.

\begin{figure}[tb]
	\includegraphics[width=0.95\textwidth]{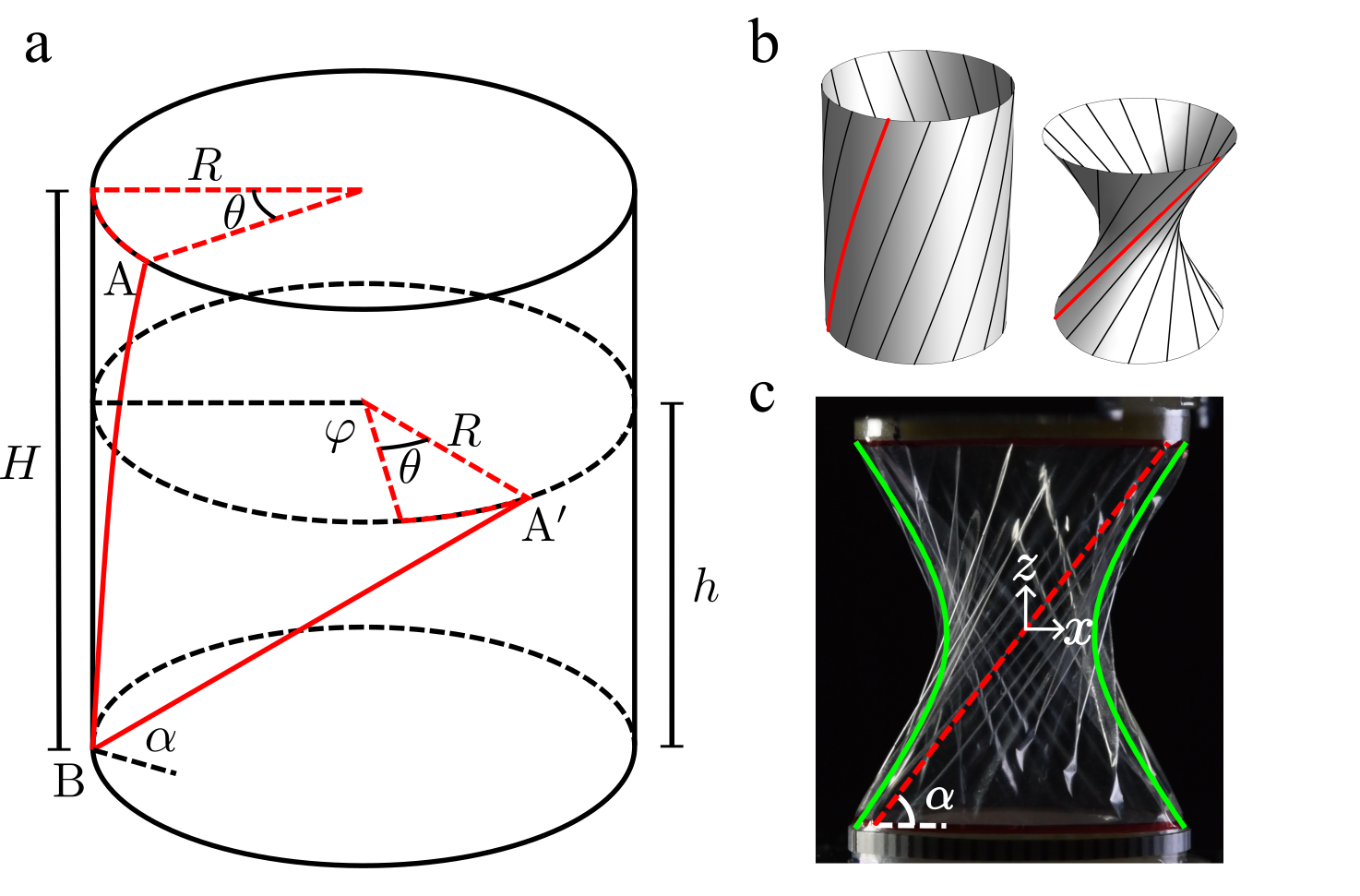}
	\caption{
	\textbf{Geometric idealization for compressing and twisting an inextensible cylindrical shell.}
	 (a) A cylindrical shell with an initial height $H$ is compressed to a final height $h$ and then twisted by $\varphi$.  In this process, a geodesic $AB$, represented by the offset angle $\theta$, is mapped onto a straight line $A'B$.  Finding the largest allowed twist $\varphi_\ell$ that does not stretch the material between any two points amounts to identifying the most constraining $\theta$ that gives the least amount of twist while preserving the length $AB=A'B$.
	 (b) An undeformed cylindrical shell and its idealized gross shape after compression and twist, showing the corresponding geodesics preserving the length. 
	 (c) 
  Shape of a locked cylinder ($\varphi=\varphi_\ell$) with $C=0.64$ and $\rho=1.42$.
  Green curves: Hyperbola given by Eq.~\ref{eq:profile} in the marked $x$-$z$ coordinate system. Red line: Asymptote of the hyperbola, which matches the wrinkle angle $\alpha$. 
	}
	\label{fig:model}
\end{figure}

\textit{Geometric model. ---} 
To understand the emergence of ordered wrinkles at a qualitative level, we imagine that the cylinder of initial 
height $H$ is constructed of many vertical ``ropes'', which are inextensible yet have zero resistance to bending. When the cylinder is compressed to a smaller height $h$, these ropes buckle to collect the extra slack along their length. This buckling of the ropes serves as a reservoir of material that allows the cylinder to then be twisted by $\varphi$, until at some crucial angle $\varphi_\ell$ the slack runs out and the ropes become taut. The same picture holds if the initial direction of the ropes is skew, forming a set of geodesics that can be characterized by an offset angle $\theta$ at the cylindrical surface (see the segment $\arc{AB}$ in Fig.~\ref{fig:model}a). As we will now show, there is a precise $\theta$ of the tilted ropes that forms the strongest constraint to the rotation. It is the geometric selection of $\theta$ that determines the locking angle $\varphi_\ell$ and the final orientation $\alpha$ of the wrinkles for the buckled configuration. 

The key geometric observation is that the initial length of the rope $\arc{AB}=\sqrt{H^2+R^2\theta^2}$, must match the final length of the rope $\overline{A’B}=\sqrt{h^2+2R^2-2R^2\cos(\theta+\varphi)}$ at the point the system becomes taut. 
Equating these two and introducing a dimensionless compression parameter and a dimensionless aspect ratio, 
\begin{align}
C&\equiv(H^2-h^2)/(2R)^2 \nonumber\\
\rho&\equiv H/(2R), \nonumber
\end{align}
we obtain $1-\theta^2/2-\cos(\theta+\varphi)=2C$. To see which $\theta$ poses the strongest constraint to rotation, we seek the value of $\theta$ for which $\varphi$ is minimized; physically this corresponds to finding the set of lines in the undeformed cylindrical shell that become taut first, as the compressed shell is gradually twisted. 
The amount of twist $\varphi_\ell$ that locks the system is thus:
\begin{align}
\varphi_\ell &\equiv \min_\theta \{\varphi(\theta)\} \nonumber\\
&=-2\sqrt{\sqrt{C}-C}+\cos^{-1}(1-2\sqrt{C}).
\label{eq:phi_l}
\end{align}
This minimum $\varphi$ occurs at $\theta_\text{m}=2\sqrt{\sqrt{C}-C}$, which identifies the material lines in the undeformed cylinder that lock the system.

To help visualize this picture, Fig.~\ref{fig:model}b is an illustration of these material lines (black helices wrapping the cylinder), and their taut counterparts in the compressed, twisted shell at $\varphi_\ell$. 
Each taut line makes an angle $\alpha$ with the horizontal. 
To obtain an expression for $\alpha$, we use the geometry in Fig.~\ref{fig:model}a to express $\alpha$ in terms of $\varphi_\ell + \theta_\text{m}$, $h$, and $R$.
Applying the definitions of the compression parameter $C$ and the aspect ratio $\rho$, we find:
\begin{align}
\alpha=\tan^{-1}\sqrt{\frac{\rho^2-C}{\sqrt{C}}}.
\label{eq:alpha}
\end{align} 

We may also use these geometric arguments to obtain the overall shape of the shell at $\varphi_\ell$ (neglecting the wrinkly undulations). 
Because each point on the locked shell belongs to one taut line, the overall shape 
is given by sweeping the straight line $\overline{A'B}$ in Fig.~\ref{fig:model}a about the axis of the cylinder. 
This surface thus generated is 
a hyperboloid of revolution \cite{hilbert21,steinhaus99}, and its side-profile 
is the hyperbola:
\begin{align}
x^2/a^2-z^2/c^2=R^2.
\label{eq:profile}
\end{align}
Remarkably, the angle $\alpha$ of the line $\overline{A'B}$ is equal to the angle of the asymptote of the side-view profile \cite{gray17}, $z = (c/a) \cdot x = \tan\alpha\cdot x$. This allows us to express the ratio $c/a$ in terms of $\rho$ and $C$, through Eq.~\ref{eq:alpha}. 
To obtain $c$ and $a$, we combine this result with Eq.~\ref{eq:profile} applied to the top boundary of the cylinder, $R^2/a^2-h^2/(4c^2) = R^2$. 
Solving these two equations gives: $a^2 = 1-\sqrt{C}$ and $c^2=(\rho^2-C)(1-\sqrt{C})/\sqrt{C}$.

As a first test of our model, we compare these predictions with the cylinder from Figs.~\ref{fig:force_torque} and \ref{fig:model}c with $C=0.64$ and $\rho=1.42$. 
Figure~\ref{fig:model}c shows the predicted hyperbolic shape, which is in reasonable agreement with the observed side-view profile, allowing for the finite-amplitude wrinkles that protrude from the hyperboloid shape, and that we neglect in the model. 
We also show the asymptote, which agrees with the orientation of the wrinkles passing through the center of the shell in the plane normal to the line of sight. 
Finally, the locking angle predicted by Eq.~\ref{eq:phi_l}, $\varphi_\ell=81^\circ$, is in reasonable agreement with the location of the rapid increase in magnitude of the force and torque curves in Fig.~\ref{fig:force_torque}b,c. 

\begin{figure}[tb]
	\includegraphics[width=0.95\textwidth]{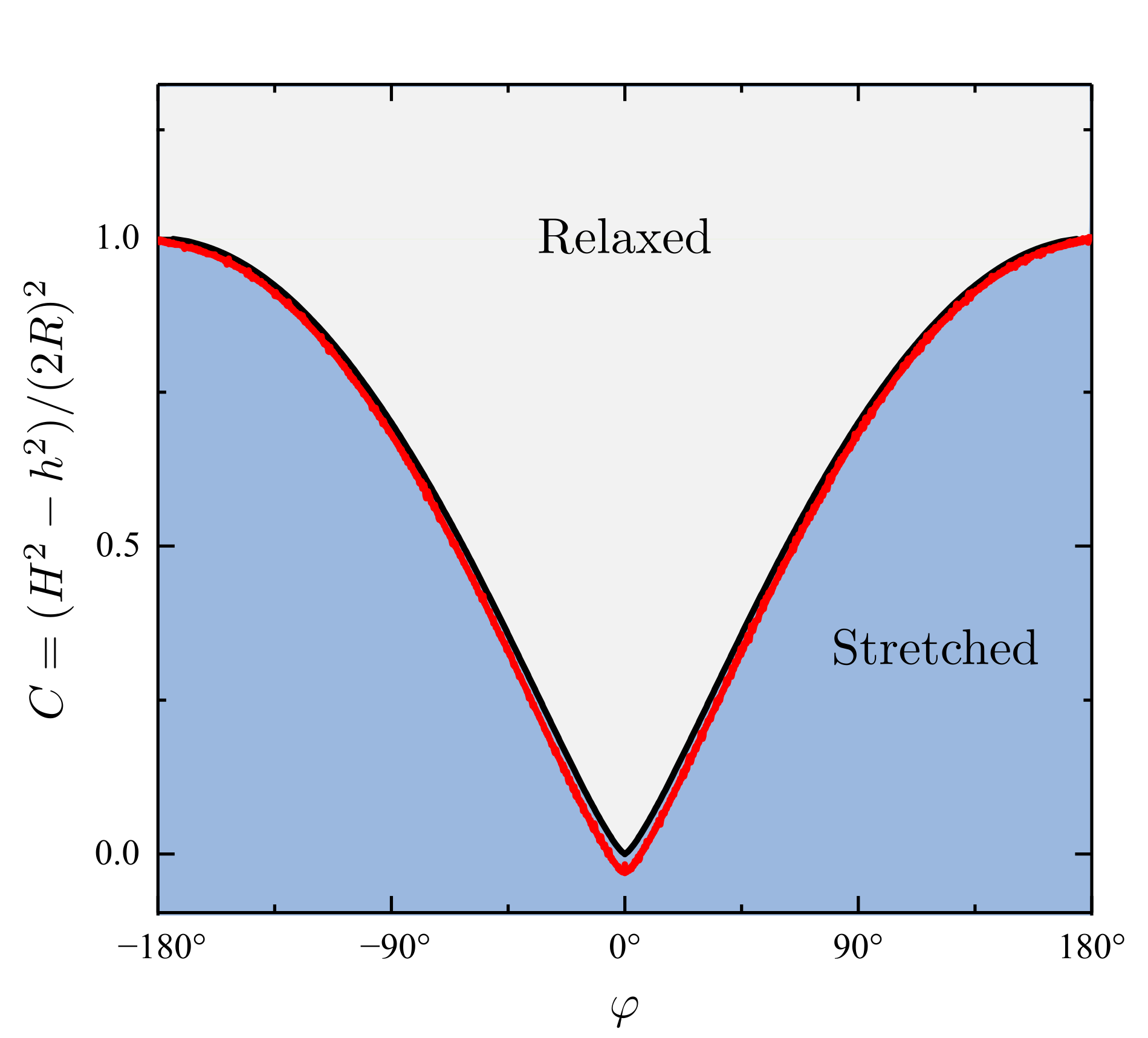}
	\caption{
	\textbf{Phase diagram of a cylindrical shell subjected to axial compression $C$ and twist $\varphi$.}
 	Modeling the system as a material that can compress but not stretch makes every point in the ``relaxed'' phase accessible (up to the horizontal line $C=\rho^2$ that corresponds to full compression with $h=0$). 
 	Our geometric model predicts the phase boundary where the system locks and becomes taut (Eq.~\ref{eq:phi_l}: black line). 
  	Red curve: Experiment with a $1.5$ $\mu$m thick shell with $R=9.5$ mm and $H=26.9$ mm,
	cycling between $\varphi=\pm180^{\circ}$ at a constant rate while applying a small lifting force $F=-0.5$ N at the top boundary. The data are in good agreement with our model that has no free parameters. The small discrepancy around $\varphi = 0$ is due to the finite stretching of the sheet; see SM.
   	}
	\label{fig:phase}
\end{figure}

\textit{Tunability. ---} A more comprehensive test of the predicted locking angle involves probing it as a function of the axial compression. 
Indeed, our geometric model predicts the phase boundary between relaxed and stretched states, and we may plot Eq.~\ref{eq:phi_l} on a phase diagram as in Fig.~\ref{fig:phase}. 
At a given axial compression, $C$, the shell is soft to rotation within the interval $|\varphi|<\varphi_\ell$. 
Figure~\ref{fig:phase} shows that the width of this interval can be tuned on demand between $0$ and $360^\circ$.

To test this picture, we designed a protocol to trace out the phase boundary in a single experiment.  
We twist the cylinder back and forth between $\pm360^{\circ}$ at a constant rate, keeping a small separation force $-0.5$ N between the top and bottom plates of the cylinder.
The exact value of this force is not important, but it must be small enough to avoid tearing the sheet, and large enough to activate relaxations of the wrinkle pattern toward the lowest-energy ordered state (see SM). 
This force causes the cylinder to gradually decrease or increase its height in response to the continuous twisting, maintaining a taut configuration.  In this process we track the twist angle $\varphi$ and the resulting compression $C$, calculated from the initial height $H$ and the adapted height $h$.  Figure~\ref{fig:phase} shows an excellent match between the 
curve traced out by the test cycle 
and the 
relation from Eq.~\ref{eq:phi_l}, 
within the working regime $|\varphi| < 180^\circ$ of our geometric model. 
This is notable given that our model does not include the bending or stretching moduli of the sheet, and it does not take into account the finite size of the wrinkles or the boundary layer where the wrinkle amplitude decays to zero at the clamped boundaries.

\begin{figure*}[tb]
	\includegraphics[width=1\textwidth]{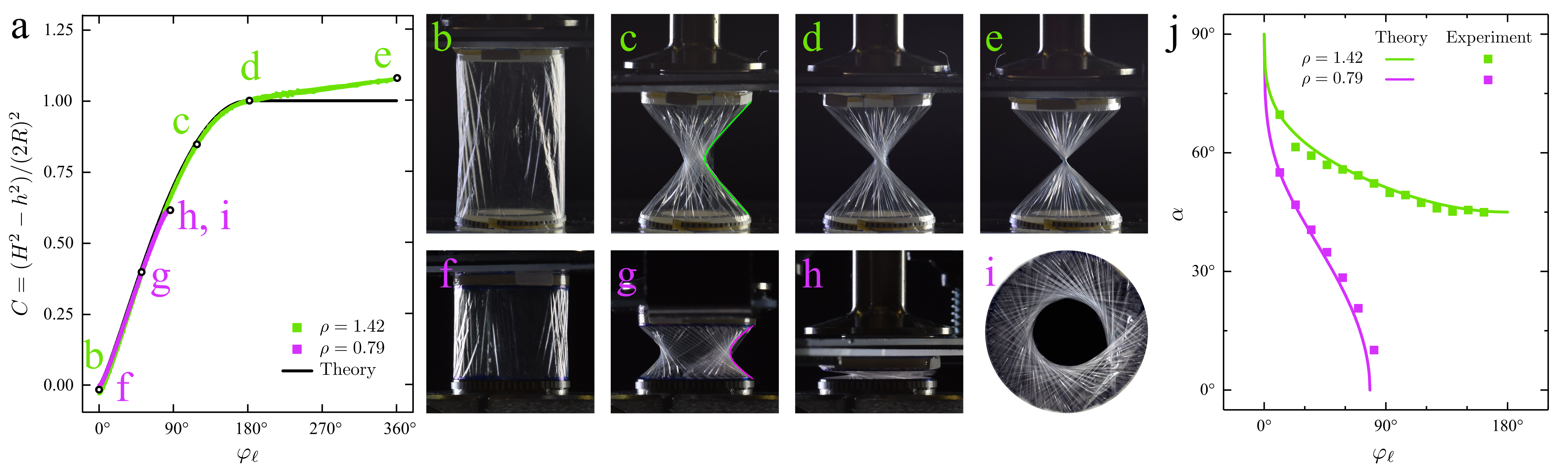}
	\caption{
	\textbf{Effects of the aspect ratio $\rho$ of the cylinder.}
	(a) Locking curves for two $1.5$ $\mu$m-thick cylindrical shells with the same radius $R=9.52$ mm but different heights, twisted from $0^{\circ}$ to $360^{\circ}$ at a constant rate while applying a small lifting force $F=-0.5$ N at the top boundary.
	Green: $H=26.9$ mm, $\rho=1.42$. 
 	Magenta: $H=15.0$ mm, $\rho=0.79$. 
	(b--e) Side-view images of the slender  shell  ($\rho=1.42$) at $\varphi=0^{\circ}$, $120^{\circ}$, $180^{\circ}$, $360^{\circ}$. 
	(f--h) Side-view images of the squat  shell ($\rho=0.79$) at $\varphi=0^{\circ}$, $52^{\circ}$, $87^{\circ}$. 
	(i) Bottom-view image of the configuration of (h).
	(j) Markers: observed angle of wrinkles with respect to the vertical direction, versus the locking angle $\varphi_\ell$, for the two cylinders. Curves: Eq.~\ref{eq:alpha} for the angle of the lines of tension.
 }
	\label{fig:aspect}
\end{figure*}

There is another striking simplicity of the behavior: 
Eq.~\ref{eq:phi_l} predicts that the 
phase diagram
is independent of the initial shape of the cylinder---$\rho$ is absent from Eq.~\ref{eq:phi_l} and the phase diagram axes (Fig.~\ref{fig:phase}). 
To further demonstrate this universality, we perform a new experiment with a cylinder of aspect ratio $\rho=0.79$ and compare it to the earlier test with $\rho=1.42$. 
Figure~\ref{fig:aspect}a shows that the curves 
followed by the slender and squat cylinders both fall along our model prediction (Eq.~\ref{eq:phi_l}). 

The fates of the two cylinders at their limits of rotation, however, are different. 
When the slender cylinder ($\rho=1.42$) is twisted toward $180^{\circ}$, a tight waist appears that contracts to a small size to form a double-cone structure (Fig.~\ref{fig:aspect}b-e). 
Indeed, at $\varphi_\ell=180^{\circ}$ ($C=1$), our geometric argument predicts that all lines of tension (represented by $\overline{A'B}$ in Fig.~\ref{fig:model}a) simultaneously pass through the center of the cylinder. 
A simple way for our model to handle this self-contact is for the lines of tension to curl around one another at the apex of the double-cone, leaving $h$ fixed even as the twist exceeds $180^{\circ}$.
This idealized behavior is represented by the horizontal line $C=1$ for $\varphi_\ell > 180^{\circ}$ in Fig.~\ref{fig:aspect}a. 
The torque is zero along this line as the lines of tension have lost their azimuthal component. 
In the experiment, the finite thickness of the shell causes a finite-sized waist to form, so that 
the dimensionless compression $C$ rises slowly for $\varphi_\ell > 180^{\circ}$, and a small torque is observed. 

On the other hand, a squat cylinder with $\rho = H/(2R) < 1$ cannot be twisted up to $\varphi = 180^{\circ}$. 
At $\varphi_{\ell}=-2\sqrt{\rho-\rho^2}+\cos^{-1}(1-2\rho) < 180^{\circ}$, $h = 0$ and the top and bottom come into contact as shown in the sequence of images in Fig.~\ref{fig:aspect}f-h for an experimental realization with $\rho=0.79$. 
This deformation results in a finite throat size, seen in the bottom view of Fig.~\ref{fig:aspect}i. 
As a consequence of this evolution, a squat cylinder never loses its locking ability. 

The orientation of the lines of stress also depends on the aspect ratio $\rho$. 
Figure~\ref{fig:aspect}j shows the measured wrinkle tilt $\alpha$ for the same two cylinders but different aspect ratios ($\rho=1.42$ and 
$0.79$) at a series of locked configurations. For the slender cylinder ($\rho>1$), the wrinkle direction plateaus to the double-cone limit at large twist. For the squat cylinder ($\rho<1$), the wrinkles become more and more skew until they lay down completely as the shell flattens into a wrinkled annulus. 
Both trends are captured accurately by our geometric prediction, Eq.~\ref{eq:alpha}.

\textit{Discussion. ---}
We have shown how a thin cylindrical shell can be manipulated to give rise to a tunable locking behavior, where the torque and force dramatically increase beyond the locking angles $\pm \varphi_\ell$. 
Beyond the locking point the material stretches, so that in the blue shaded region of Fig.~\ref{fig:phase} the system acts as an extensional or torsional spring \cite{Tait96}.
Our geometric arguments show how $\varphi_\ell$ can be set on demand over a wide range---anywhere from 0 to $180^{\circ}$---and our results apply to sufficiently thin cylindrical shells of any aspect ratio. 
This behavior is amenable to applications, as the system is lightweight and relies on a low-cost film that can be sourced from roll-to-roll processing. 

At a twist angle of $180^{\circ}$, our purely geometric arguments show that the sheet must make contact with itself. As we approach this angle from below, the waist of the sheet approaches zero radius, wrinkles become more prevalent, and the sheet’s finite bending modulus and thickness play a role, so that the experimental data begin to deviate from our model. Eventually a tight bundle forms where the material twists around itself in a neck that joins two wrinkled cones---reminiscent of a twisted candy wrapper, sausage casing, party balloon~\cite{Cheng21} or towel. An understanding of this bundle must take into account the finite thickness of the film, the friction of self-contact, and the geometric problem of packing the sheet into a small cross-section~\cite{Chopin22}. 

Our results suggest a fascinating landscape of configurations for the sheet. 
In our idealized model, each point on the phase boundary corresponds to a unique configuration of the sheet, where uniform wrinkles orient at an angle $\alpha$ around a hyperbolic profile. This is reflected in our experiments as a robust and repeatable ordered response at the phase boundary, although the precise placement of wrinkles can differ in each realization. 
Slightly away from the phase boundary, the number of possible states expands dramatically: even though the wrinkles still have a dominant orientation, some are buckled and some have other directions. Finally, deep into the ``relaxed'' phase (the gray shaded region in Fig.~\ref{fig:phase}), the sheet can be in a vast number of metastable states (Fig.~\ref{fig:force_torque}a, left image), in contrast to the relatively small set of ``bottleneck'' states along the boundary. 
We note that this evolution from smooth wrinkles to sharp crumpled morphologies evades an existing empirical framework for such a transition \cite{King12,Timounay20}, as here the crumples arise when there are no tensional loads on the boundaries of the shell. 

Equally striking is the difference among the possible paths from an arbitrary $\varphi_\ell$ with one set of ordered wrinkles, to $-\varphi_\ell$ with a mutually incompatible set. Uniquely, when one travels along the phase boundary the wrinkles deform smoothly as $\alpha$ changes, and they vanish momentarily at $\varphi_\ell = 0$. All other paths traverse the bulk, compressing and buckling the initial wrinkles, and then straightening out the resulting network via many snap-through events---a noisy, messy, and highly path-dependent~\cite{Shohat22} journey that showcases the essential glassiness of a crumpled sheet.

\begin{acknowledgements}
We thank Ian Tobasco and Helen Ansell for informative discussions. 
Funding support from NSF-DMR-CAREER-1654102 (P.D., M.H., J.D.P.) is gratefully acknowledged. 
\end{acknowledgements}


\begin{thebibliography}{21}%
\makeatletter
\providecommand \@ifxundefined [1]{%
 \@ifx{#1\undefined}
}%
\providecommand \@ifnum [1]{%
 \ifnum #1\expandafter \@firstoftwo
 \else \expandafter \@secondoftwo
 \fi
}%
\providecommand \@ifx [1]{%
 \ifx #1\expandafter \@firstoftwo
 \else \expandafter \@secondoftwo
 \fi
}%
\providecommand \natexlab [1]{#1}%
\providecommand \enquote  [1]{``#1''}%
\providecommand \bibnamefont  [1]{#1}%
\providecommand \bibfnamefont [1]{#1}%
\providecommand \citenamefont [1]{#1}%
\providecommand \href@noop [0]{\@secondoftwo}%
\providecommand \href [0]{\begingroup \@sanitize@url \@href}%
\providecommand \@href[1]{\@@startlink{#1}\@@href}%
\providecommand \@@href[1]{\endgroup#1\@@endlink}%
\providecommand \@sanitize@url [0]{\catcode `\\12\catcode `\$12\catcode
  `\&12\catcode `\#12\catcode `\^12\catcode `\_12\catcode `\%12\relax}%
\providecommand \@@startlink[1]{}%
\providecommand \@@endlink[0]{}%
\providecommand \url  [0]{\begingroup\@sanitize@url \@url }%
\providecommand \@url [1]{\endgroup\@href {#1}{\urlprefix }}%
\providecommand \urlprefix  [0]{URL }%
\providecommand \Eprint [0]{\href }%
\providecommand \doibase [0]{https://doi.org/}%
\providecommand \selectlanguage [0]{\@gobble}%
\providecommand \bibinfo  [0]{\@secondoftwo}%
\providecommand \bibfield  [0]{\@secondoftwo}%
\providecommand \translation [1]{[#1]}%
\providecommand \BibitemOpen [0]{}%
\providecommand \bibitemStop [0]{}%
\providecommand \bibitemNoStop [0]{.\EOS\space}%
\providecommand \EOS [0]{\spacefactor3000\relax}%
\providecommand \BibitemShut  [1]{\csname bibitem#1\endcsname}%
\let\auto@bib@innerbib\@empty
\bibitem [{\citenamefont {Lennon}\ and\ \citenamefont
  {Pellegrino}(2005)}]{Lennon05}%
  \BibitemOpen
  \bibfield  {author} {\bibinfo {author} {\bibfnamefont {A.}~\bibnamefont
  {Lennon}}\ and\ \bibinfo {author} {\bibfnamefont {S.}~\bibnamefont
  {Pellegrino}},\ }in\ \href@noop {} {\emph {\bibinfo {booktitle} {Proc. Euro.
  Conf. Space. Struct. Mater. Mech. Testing}}},\ \bibinfo {editor} {edited by\
  \bibinfo {editor} {\bibfnamefont {K.}~\bibnamefont {Fletcher}}},\ \bibinfo
  {organization} {Noordwijk, The Netherlands, May 10-12, p. 101}\ (\bibinfo
  {publisher} {ESA},\ \bibinfo {address} {Noordwijk, The Netherlands},\
  \bibinfo {year} {2005})\BibitemShut {NoStop}%
\bibitem [{\citenamefont {Paulsen}(2019)}]{Paulsen19}%
  \BibitemOpen
  \bibfield  {author} {\bibinfo {author} {\bibfnamefont {J.~D.}\ \bibnamefont
  {Paulsen}},\ }\bibfield  {title} {\bibinfo {title} {Wrapping liquids, solids,
  and gases in thin sheets},\ }\href
  {https://doi.org/10.1146/annurev-conmatphys-031218-013533} {\bibfield
  {journal} {\bibinfo  {journal} {Annual Review of Condensed Matter Physics}\
  }\textbf {\bibinfo {volume} {10}},\ \bibinfo {pages} {431} (\bibinfo {year}
  {2019})}\BibitemShut {NoStop}%
\bibitem [{\citenamefont {Elettro}\ \emph {et~al.}(2016)\citenamefont
  {Elettro}, \citenamefont {Neukirch}, \citenamefont {Vollrath},\ and\
  \citenamefont {Antkowiak}}]{Elettro16}%
  \BibitemOpen
  \bibfield  {author} {\bibinfo {author} {\bibfnamefont {H.}~\bibnamefont
  {Elettro}}, \bibinfo {author} {\bibfnamefont {S.}~\bibnamefont {Neukirch}},
  \bibinfo {author} {\bibfnamefont {F.}~\bibnamefont {Vollrath}},\ and\
  \bibinfo {author} {\bibfnamefont {A.}~\bibnamefont {Antkowiak}},\ }\bibfield
  {title} {\bibinfo {title} {In-drop capillary spooling of spider capture
  thread inspires hybrid fibers with mixed solid?liquid mechanical
  properties},\ }\href {https://doi.org/10.1073/pnas.1602451113} {\bibfield
  {journal} {\bibinfo  {journal} {Proceedings of the National Academy of
  Sciences}\ }\textbf {\bibinfo {volume} {113}},\ \bibinfo {pages} {6143}
  (\bibinfo {year} {2016})}\BibitemShut {NoStop}%
\bibitem [{\citenamefont {Grandgeorge}\ \emph {et~al.}(2018)\citenamefont
  {Grandgeorge}, \citenamefont {Krins}, \citenamefont {Hourlier-Fargette},
  \citenamefont {Laberty-Robert}, \citenamefont {Neukirch},\ and\ \citenamefont
  {Antkowiak}}]{Grandgeorge18}%
  \BibitemOpen
  \bibfield  {author} {\bibinfo {author} {\bibfnamefont {P.}~\bibnamefont
  {Grandgeorge}}, \bibinfo {author} {\bibfnamefont {N.}~\bibnamefont {Krins}},
  \bibinfo {author} {\bibfnamefont {A.}~\bibnamefont {Hourlier-Fargette}},
  \bibinfo {author} {\bibfnamefont {C.}~\bibnamefont {Laberty-Robert}},
  \bibinfo {author} {\bibfnamefont {S.}~\bibnamefont {Neukirch}},\ and\
  \bibinfo {author} {\bibfnamefont {A.}~\bibnamefont {Antkowiak}},\ }\bibfield
  {title} {\bibinfo {title} {Capillarity-induced folds fuel extreme shape
  changes in thin wicked membranes},\ }\href
  {https://doi.org/10.1126/science.aaq0677} {\bibfield  {journal} {\bibinfo
  {journal} {Science}\ }\textbf {\bibinfo {volume} {360}},\ \bibinfo {pages}
  {296} (\bibinfo {year} {2018})}\BibitemShut {NoStop}%
\bibitem [{\citenamefont {Fung}(1967)}]{Fung67}%
  \BibitemOpen
  \bibfield  {author} {\bibinfo {author} {\bibfnamefont {Y.}~\bibnamefont
  {Fung}},\ }\bibfield  {title} {\bibinfo {title} {Elasticity of soft tissues
  in simple elongation},\ }\href
  {https://doi.org/10.1152/ajplegacy.1967.213.6.1532} {\bibfield  {journal}
  {\bibinfo  {journal} {American Journal of Physiology-Legacy Content}\
  }\textbf {\bibinfo {volume} {213}},\ \bibinfo {pages} {1532} (\bibinfo {year}
  {1967})}\BibitemShut {NoStop}%
\bibitem [{\citenamefont {Tobasco}\ \emph {et~al.}(2022)\citenamefont
  {Tobasco}, \citenamefont {Timounay}, \citenamefont {Todorova}, \citenamefont
  {Leggat}, \citenamefont {Paulsen},\ and\ \citenamefont
  {Katifori}}]{Tobasco22}%
  \BibitemOpen
  \bibfield  {author} {\bibinfo {author} {\bibfnamefont {I.}~\bibnamefont
  {Tobasco}}, \bibinfo {author} {\bibfnamefont {Y.}~\bibnamefont {Timounay}},
  \bibinfo {author} {\bibfnamefont {D.}~\bibnamefont {Todorova}}, \bibinfo
  {author} {\bibfnamefont {G.~C.}\ \bibnamefont {Leggat}}, \bibinfo {author}
  {\bibfnamefont {J.~D.}\ \bibnamefont {Paulsen}},\ and\ \bibinfo {author}
  {\bibfnamefont {E.}~\bibnamefont {Katifori}},\ }\bibfield  {title} {\bibinfo
  {title} {Exact solutions for the wrinkle patterns of confined elastic
  shells},\ }\href {https://doi.org/10.1038/s41567-022-01672-2} {\bibfield
  {journal} {\bibinfo  {journal} {Nature Physics}\ }\textbf {\bibinfo {volume}
  {18}},\ \bibinfo {pages} {1099} (\bibinfo {year} {2022})}\BibitemShut
  {NoStop}%
\bibitem [{\citenamefont {Prager}(1957)}]{Prager57}%
  \BibitemOpen
  \bibfield  {author} {\bibinfo {author} {\bibfnamefont {W.}~\bibnamefont
  {Prager}},\ }\bibfield  {title} {\bibinfo {title} {On ideal locking
  materials},\ }\href {https://doi.org/10.1122/1.548818} {\bibfield  {journal}
  {\bibinfo  {journal} {Transactions of the Society of Rheology}\ }\textbf
  {\bibinfo {volume} {1}},\ \bibinfo {pages} {169} (\bibinfo {year}
  {1957})}\BibitemShut {NoStop}%
\bibitem [{\citenamefont {Prager}(1969)}]{Prager69}%
  \BibitemOpen
  \bibfield  {author} {\bibinfo {author} {\bibfnamefont {W.}~\bibnamefont
  {Prager}},\ }\bibfield  {title} {\bibinfo {title} {On the formulation of
  constitutive equations for living soft tissues},\ }\href@noop {} {\bibfield
  {journal} {\bibinfo  {journal} {Quarterly of Applied Mathematics}\ }\textbf
  {\bibinfo {volume} {27}},\ \bibinfo {pages} {128} (\bibinfo {year}
  {1969})}\BibitemShut {NoStop}%
\bibitem [{\citenamefont {Wagner}(1929)}]{Wagner29}%
  \BibitemOpen
  \bibfield  {author} {\bibinfo {author} {\bibfnamefont {H.}~\bibnamefont
  {Wagner}},\ }\bibfield  {title} {\bibinfo {title} {Ebene blechwandtrager mit
  sehr dunnem stegblech (metal beams with very thin webs)},\ }\href@noop {}
  {\bibfield  {journal} {\bibinfo  {journal} {Zeitschrift fur Flugtechnik und
  Motorloftschiffahr}\ }\textbf {\bibinfo {volume} {20}} (\bibinfo {year}
  {1929})}\BibitemShut {NoStop}%
\bibitem [{\citenamefont {Pipkin}(1986)}]{Pipkin86}%
  \BibitemOpen
  \bibfield  {author} {\bibinfo {author} {\bibfnamefont {A.~C.}\ \bibnamefont
  {Pipkin}},\ }\bibfield  {title} {\bibinfo {title} {{The Relaxed Energy
  Density for Isotropic Elastic Membranes}},\ }\href
  {https://doi.org/10.1093/imamat/36.1.85} {\bibfield  {journal} {\bibinfo
  {journal} {IMA Journal of Applied Mathematics}\ }\textbf {\bibinfo {volume}
  {36}},\ \bibinfo {pages} {85} (\bibinfo {year} {1986})}\BibitemShut {NoStop}%
\bibitem [{\citenamefont {{Mansfield}}(1989)}]{Mansfield89}%
  \BibitemOpen
  \bibfield  {author} {\bibinfo {author} {\bibfnamefont {E.~H.}\ \bibnamefont
  {{Mansfield}}},\ }\href@noop {} {\emph {\bibinfo {title} {{The Bending and
  Stretching of Plates}}}}\ (\bibinfo {year} {1989})\BibitemShut {NoStop}%
\bibitem [{\citenamefont {Steigmann}\ and\ \citenamefont
  {Green}(1990)}]{Steigmann90}%
  \BibitemOpen
  \bibfield  {author} {\bibinfo {author} {\bibfnamefont {D.~J.}\ \bibnamefont
  {Steigmann}}\ and\ \bibinfo {author} {\bibfnamefont {A.~E.}\ \bibnamefont
  {Green}},\ }\bibfield  {title} {\bibinfo {title} {Tension-field theory},\
  }\href {https://doi.org/10.1098/rspa.1990.0055} {\bibfield  {journal}
  {\bibinfo  {journal} {Proceedings of the Royal Society of London. A.
  Mathematical and Physical Sciences}\ }\textbf {\bibinfo {volume} {429}},\
  \bibinfo {pages} {141} (\bibinfo {year} {1990})}\BibitemShut {NoStop}%
\bibitem [{\citenamefont {Hilbert}\ and\ \citenamefont
  {Cohn-Vossen}(2021)}]{hilbert21}%
  \BibitemOpen
  \bibfield  {author} {\bibinfo {author} {\bibfnamefont {D.}~\bibnamefont
  {Hilbert}}\ and\ \bibinfo {author} {\bibfnamefont {S.}~\bibnamefont
  {Cohn-Vossen}},\ }\href@noop {} {\emph {\bibinfo {title} {Geometry and the
  Imagination}}},\ Vol.~\bibinfo {volume} {87}\ (\bibinfo  {publisher}
  {American Mathematical Soc.},\ \bibinfo {year} {2021})\BibitemShut {NoStop}%
\bibitem [{\citenamefont {Steinhaus}(1999)}]{steinhaus99}%
  \BibitemOpen
  \bibfield  {author} {\bibinfo {author} {\bibfnamefont {H.}~\bibnamefont
  {Steinhaus}},\ }\href@noop {} {\emph {\bibinfo {title} {Mathematical
  snapshots}}}\ (\bibinfo  {publisher} {Courier Corporation},\ \bibinfo {year}
  {1999})\BibitemShut {NoStop}%
\bibitem [{\citenamefont {Gray}\ \emph {et~al.}(2017)\citenamefont {Gray},
  \citenamefont {Abbena},\ and\ \citenamefont {Salamon}}]{gray17}%
  \BibitemOpen
  \bibfield  {author} {\bibinfo {author} {\bibfnamefont {A.}~\bibnamefont
  {Gray}}, \bibinfo {author} {\bibfnamefont {E.}~\bibnamefont {Abbena}},\ and\
  \bibinfo {author} {\bibfnamefont {S.}~\bibnamefont {Salamon}},\ }\href@noop
  {} {\emph {\bibinfo {title} {Modern differential geometry of curves and
  surfaces with Mathematica{\textregistered}}}}\ (\bibinfo  {publisher}
  {Chapman and Hall/CRC},\ \bibinfo {year} {2017})\BibitemShut {NoStop}%
\bibitem [{\citenamefont {Tait}\ \emph {et~al.}(1996)\citenamefont {Tait},
  \citenamefont {Steigmann},\ and\ \citenamefont {Zhong}}]{Tait96}%
  \BibitemOpen
  \bibfield  {author} {\bibinfo {author} {\bibfnamefont {R.}~\bibnamefont
  {Tait}}, \bibinfo {author} {\bibfnamefont {D.}~\bibnamefont {Steigmann}},\
  and\ \bibinfo {author} {\bibfnamefont {J.}~\bibnamefont {Zhong}},\ }\bibfield
   {title} {\bibinfo {title} {Finite twist and extension of a cylindrical
  elastic membrane},\ }\href@noop {} {\bibfield  {journal} {\bibinfo  {journal}
  {Acta mechanica}\ }\textbf {\bibinfo {volume} {117}},\ \bibinfo {pages} {129}
  (\bibinfo {year} {1996})}\BibitemShut {NoStop}%
\bibitem [{\citenamefont {Cheng}\ \emph {et~al.}(2021)\citenamefont {Cheng},
  \citenamefont {Hsieh}, \citenamefont {Tsai},\ and\ \citenamefont
  {Hong}}]{Cheng21}%
  \BibitemOpen
  \bibfield  {author} {\bibinfo {author} {\bibfnamefont {Y.-C.}\ \bibnamefont
  {Cheng}}, \bibinfo {author} {\bibfnamefont {T.-H.}\ \bibnamefont {Hsieh}},
  \bibinfo {author} {\bibfnamefont {J.-C.}\ \bibnamefont {Tsai}},\ and\
  \bibinfo {author} {\bibfnamefont {T.-M.}\ \bibnamefont {Hong}},\ }\bibfield
  {title} {\bibinfo {title} {Phase diagram and snap-off transition for twisted
  party balloons},\ }\href {https://doi.org/10.1103/PhysRevE.104.045004}
  {\bibfield  {journal} {\bibinfo  {journal} {Phys. Rev. E}\ }\textbf {\bibinfo
  {volume} {104}},\ \bibinfo {pages} {045004} (\bibinfo {year}
  {2021})}\BibitemShut {NoStop}%
\bibitem [{\citenamefont {Chopin}\ and\ \citenamefont
  {Kudrolli}(2022)}]{Chopin22}%
  \BibitemOpen
  \bibfield  {author} {\bibinfo {author} {\bibfnamefont {J.}~\bibnamefont
  {Chopin}}\ and\ \bibinfo {author} {\bibfnamefont {A.}~\bibnamefont
  {Kudrolli}},\ }\bibfield  {title} {\bibinfo {title} {Tensional twist-folding
  of sheets into multilayered scrolled yarns},\ }\href
  {https://doi.org/10.1126/sciadv.abi8818} {\bibfield  {journal} {\bibinfo
  {journal} {Science Advances}\ }\textbf {\bibinfo {volume} {8}},\ \bibinfo
  {pages} {eabi8818} (\bibinfo {year} {2022})}\BibitemShut {NoStop}%
\bibitem [{\citenamefont {King}\ \emph {et~al.}(2012)\citenamefont {King},
  \citenamefont {Schroll}, \citenamefont {Davidovitch},\ and\ \citenamefont
  {Menon}}]{King12}%
  \BibitemOpen
  \bibfield  {author} {\bibinfo {author} {\bibfnamefont {H.}~\bibnamefont
  {King}}, \bibinfo {author} {\bibfnamefont {R.~D.}\ \bibnamefont {Schroll}},
  \bibinfo {author} {\bibfnamefont {B.}~\bibnamefont {Davidovitch}},\ and\
  \bibinfo {author} {\bibfnamefont {N.}~\bibnamefont {Menon}},\ }\bibfield
  {title} {\bibinfo {title} {Elastic sheet on a liquid drop reveals wrinkling
  and crumpling as distinct symmetry-breaking instabilities},\ }\href
  {https://doi.org/10.1073/pnas.1201201109} {\bibfield  {journal} {\bibinfo
  {journal} {PNAS}\ }\textbf {\bibinfo {volume} {109}},\ \bibinfo {pages}
  {9716} (\bibinfo {year} {2012})}\BibitemShut {NoStop}%
\bibitem [{\citenamefont {Timounay}\ \emph {et~al.}(2020)\citenamefont
  {Timounay}, \citenamefont {De}, \citenamefont {Stelzel}, \citenamefont
  {Schrecengost}, \citenamefont {Ripp},\ and\ \citenamefont
  {Paulsen}}]{Timounay20}%
  \BibitemOpen
  \bibfield  {author} {\bibinfo {author} {\bibfnamefont {Y.}~\bibnamefont
  {Timounay}}, \bibinfo {author} {\bibfnamefont {R.}~\bibnamefont {De}},
  \bibinfo {author} {\bibfnamefont {J.~L.}\ \bibnamefont {Stelzel}}, \bibinfo
  {author} {\bibfnamefont {Z.~S.}\ \bibnamefont {Schrecengost}}, \bibinfo
  {author} {\bibfnamefont {M.~M.}\ \bibnamefont {Ripp}},\ and\ \bibinfo
  {author} {\bibfnamefont {J.~D.}\ \bibnamefont {Paulsen}},\ }\bibfield
  {title} {\bibinfo {title} {Crumples as a generic stress-focusing instability
  in confined sheets},\ }\href {https://doi.org/10.1103/PhysRevX.10.021008}
  {\bibfield  {journal} {\bibinfo  {journal} {Phys. Rev. X}\ }\textbf {\bibinfo
  {volume} {10}},\ \bibinfo {pages} {021008} (\bibinfo {year}
  {2020})}\BibitemShut {NoStop}%
\bibitem [{\citenamefont {Shohat}\ \emph {et~al.}(2022)\citenamefont {Shohat},
  \citenamefont {Hexner},\ and\ \citenamefont {Lahini}}]{Shohat22}%
  \BibitemOpen
  \bibfield  {author} {\bibinfo {author} {\bibfnamefont {D.}~\bibnamefont
  {Shohat}}, \bibinfo {author} {\bibfnamefont {D.}~\bibnamefont {Hexner}},\
  and\ \bibinfo {author} {\bibfnamefont {Y.}~\bibnamefont {Lahini}},\
  }\bibfield  {title} {\bibinfo {title} {Memory from coupled instabilities in
  unfolded crumpled sheets},\ }\href {https://doi.org/10.1073/pnas.2200028119}
  {\bibfield  {journal} {\bibinfo  {journal} {Proceedings of the National
  Academy of Sciences}\ }\textbf {\bibinfo {volume} {119}},\ \bibinfo {pages}
  {e2200028119} (\bibinfo {year} {2022})}\BibitemShut {NoStop}%
\end{thebibliography}
%

\end{document}